\documentclass{ws-ijmpb}

\usepackage{ bbold }

\begin{document}

\markboth{Authors' Names}
{Instructions for Typing Manuscripts (Paper's Title)}

%
\catchline{}{}{}{}{}
%

\title{Rashba controlled two-electron spin-charge qubits \\as building blocks of a quantum computer}

\author{AMBRO\v Z KREGAR
}

\address{Institute of Physical and Theoretical Chemistry, Graz University of Technology, Stremayrgasse 9\\
A-8010 Graz, Austria \& \\
Faculty of Mechanical Engineering, University of Ljubljana, A\v sker\v ceva 6\\
1000 Ljubljana, Slovenia
\\
ambroz.kregar@fs.uni-lj.si
}

\author{ANTON RAM\v SAK}

\address{Faculty of Mathematics and Physics, University of Ljubljana, Ljubljana, Slovenia\\
J. Stefan Institute, Ljubljana, Slovenia\\
anton.ramsak@fmf.uni-lj.si}

\maketitle

\begin{history}
\end{history}

\begin{abstract}
The spin-sensitive charge oscillation, controlled by an external magnetic field, was recently proposed as a mechanism of transformations of qubits, defined as two-electron spin-charge Wannier molecules in a square quantum dot\cite{Bayat2013}. The paper expands this idea by including the effects of Rashba type spin-orbit coupling. The problem is studied theoretically by mapping the system to an analytic effective Hamiltonian for 8 low energy states, comprising singlet and triplet on each dot diagonal. The validity of the mapping is confirmed by comparing the energy and spin of full and mapped system, and also by the reproduction of charge-oscillation dynamics in the presence of magnetic flux. The newly introduced Rashba coupling significantly enriches the system dynamics, affecting the magnitude of charge oscillations and allowing the controlled transitions between singlet and triplet states due to the spin rotations, induced by spin-orbit coupling. The results indicate the possibility for use of the studied system for quantum information processing, while possible extensions of the system to serve as a qubit in a universal quantum computer, fulfilling all five DiVincenzo criteria, are also discussed.
\end{abstract}


\section{Introduction}
\label{sec:6_1_intro}

The use of quantum phenomena to increase the speed and efficiency of computation was foreseen by Richard Feynman nearly 40 years ago\cite{Feynman1982}. In the following years, many physical systems have been proposed in which the universal quantum computer could be implemented. Among the most promising and researched is the possibility of quantum computing in silicon-based devices, similar to semi-conducting transistors to currently used in classic computers\cite{Jazaeri2019}. Several implementations of qubits in semiconductor devices have recently been proposed and even experimentally tested, with qubit states based on nuclear spin of donor atom\cite{Kane1998}, single-electron spin\cite{Pla2012} or hole spin\cite{Maurand2016}.

One of interesting theoretically proposed qubit systems is a polygonal mesoscopic quantum dot, occupied by two electrons. For a dot of sufficient size, the Coulomb repulsion between electrons overcomes the kinetic energy, resulting in the formation of localized peaks in charge density. As an analogy to the "Wigner crystal"\cite{Wigner1934}, an ordered state of electrons in bulk material due to electrostatic repulsion, the described state of electrons in a dot has been named "Wigner molecule"\cite{Jauregui1993}. This charge separation provides an additional degree of freedom which might potentially be exploited for quantum information manipulation. It was shown in several papers\cite{Bayat2013,Jefferson2002,Coello2010} that the spin properties of charge-separated Wigner molecules affect the time dynamics of state evolution: in the absence of magnetic field, the singlet state's charge will oscillate while the triplet state will be stationary. The dynamics can be reversed by the application of magnetic flux through the dot. If singlet and triplet states are regarded as qubit states $\left| 0 \right\rangle$ and $\left| 1 \right\rangle$, this devices enables a simple, controlled qubit transformation\cite{Bayat2013}.

The goal of this paper is to explore how this kind of qubit manipulation device could be further enhanced by exploiting the Rashba spin-orbit coupling\cite{Rashba1960}, which forces the spin of moving electron to oscillate\cite{Datta1990}. Several proposals have been made how this effect could be used to manipulate the state of a qubit, defined as a spin-charge state of the electron, by changing its position using electric gates\cite{Ramsak2018}. This can be done by either moving the electron along the line by external potential\cite{Cadez2013,Cadez2014} or by moving it around the ring\cite{Ramsak2018,Kregar2015a,Kregar2016,Kregar2019}. It is, therefore, speculated that the addition of the Rashba coupling to the two-electron square quantum dot model of qubit will result in additional possibilities of qubit transformations, controlled by  the strength of the Rashba coupling using external voltage gates\cite{Nitta1997,Schapers1998}. The precise tuning of driving, however, will be influenced also by phonon-mediated instabilities in molecular systems with phonon assisted potential barriers\cite{mravlje06,mravlje08} or  the noise due to the electron-electron interaction\cite{rejec00,giavaras06,jefferson06}. The effects of white gate noise can in some cases be performed analytically\cite{ulcakar17}.

The effect of the Rashba coupling on two-electron quantum dot state will be studied by first numerically calculating the charge, spin and energy properties of the low-energy manifold of states in the dot. The system will then be mapped to the states of two-electrons on a 4-site quantum ring using the formalism recently developed by us\cite{Kregar2019}, producing a simple analytical $8 \times 8$ Hamiltonian. The effects of the Rashba interaction on oscillations of spin-charge states is highlighted by the study of hopping terms in Hamiltonian at different values of magnetic flux. The results show a strong effect of the Rashba coupling on both charge oscillation frequency and accompanied spin rotations, indicating a potential use of proposed devices for controlled manipulation of spin-charge based qubit states.

\section{Eigenstates and eigenenergies of two-electron square quantum dot.}
\label{sec:6_2_effham}

\begin{figure}[htbp] 
\centering 
\includegraphics[scale=1]{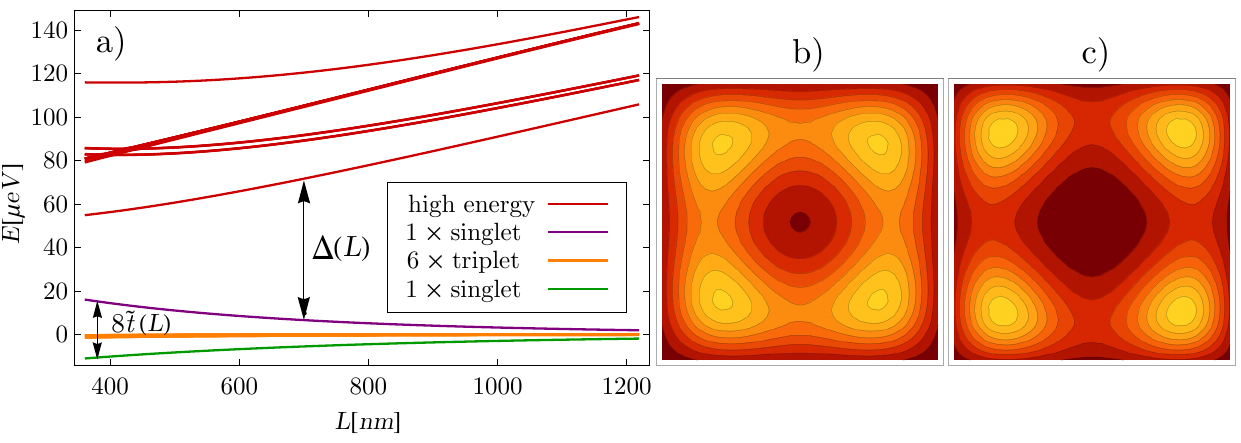} 
\caption{a) Energy levels of a system of two electrons in a square quantum dot as a function of dot size $L$. The gap $\Delta$ between 8-fold low-energy subspace and high-energy states increases with increasing $L$, while the splitting between low-energy singlet and triplet states decreases. The calculations were done for GaAs with $m=0.067\,m_0$ and $\varepsilon = 10.9$. b) Charge distribution for lowest-energy singlet eigenstate in a square quantum dot in the absence of the Rashba coupling and magnetic flux for a quantum dot of size $L=400\,\text{nm}$. c) Same as b), but for a quantum dot of size $L=800\,\text{nm}$.}
\label{fig:fig1} 
\end{figure}

The work by Creffield in 1999 showed that for sufficiently large square quantum dot, populated by two electrons, the charge separation will occur in the dot, resulting in the formation of the Wigner molecule\cite{Creffield1999}. To explore the effect of the Rashba coupling on energy, charge and spin of states in such a system, the eigenstates of the system are calculated numerically on a square grid of $16 \times 16$ sites with hard-wall boundary conditions. The two-electron Rashba Hamiltonian\cite{Liu2010}
\begin{equation}
    H = \sum_{i=1,2} \left[ \frac{1}{2 m} \left(\vec{p}_i - e \vec{A} \right)^2 + \alpha_R \vec{\sigma}_i \cdot \vec{e}_z \times \left(\vec{p}_i - e \vec{A} \right) \right] + \frac{e^2}{4 \pi \epsilon \epsilon_0 \left| \vec{r}_1 - \vec{r}_2 \right|}
\end{equation}
of the two-dimensional system is rewritten in a discrete form by the substitution of derivatives with finite differences. The Coulomb interaction is used to describe the repulsion between electrons, with values of permittivity and effective mass of the electron taken for GaAs: $\varepsilon = 10.9$ and $m = 0.067\,m_0$\cite{Creffield1999}. The eigenstates and eigenvalues of the Hamiltonian are obtained using the Lanczos algorithm. The calculations were done for dots of approximately the same sizes as those used in Refs.~\refcite{Bayat2013,Jefferson2002} and \refcite{Coello2010}, that is with the sides $L$ of several hundred nanometers. The energy spectrum of lowest energy eigenstates is shown in Fig. \ref{fig:fig1} a) as a function of $L$. We see that with increasing size of the dot, the gap between 8-fold low-energy manifold and high-energy states is increasing. Furthermore, the charge of electrons is also increasingly localized in the corners of the dot when the size increases, which is shown in Fig. \ref{fig:fig1} b) and c), for the dots of sides  $L=400\,\text{nm}$ (b) and $800\,\text{nm}$ (c) in the absence of the Rashba coupling and magnetic flux.

The 8-fold degeneracy of the ground-state manifold and strong localization of charge density on 4 sites indicate that the low-energy subspace of Hamiltonian can be effectively described as a 4-site quantum ring with 2 electrons, located on opposite sites of the ring. This assumption is verified by constructing an effective two-electron Hamiltonian for a 4-site ring in the presence of both the Rashba coupling and magnetic flux, find its eigenstates and verify that the charge distribution, energy spectra and spin properties of both square quantum dot and ring system are the same.

\section{Mapping of the system on the two-electron on 4-site quantum ring.}

The Rashba Hamiltonian of two electrons on a 4-site ring is expressed in terms of second-quantization operators $d^\dagger_{n s}$, creating localized single-electron states in the corners of the dot, labelled by $n=0,1,2,3$, with pseudo-spin $s = \pm 1/2$\cite{Kregar2019},
\begin{align}
\label{eq:6_1_H0}
H =& {H}_{kin} + {H}_C = \sum_{n,s} \left(- t_s d_{n+1,s}^\dagger d_{n s} - t^*_s d_{n-1,s}^\dagger d_{n s}\right) + \sum_{n_1,n_2} U_{n_1,n_2} n_{n_1} n_{n_2}.
\end{align}
The kinetic part of Hamiltonian $H_{kin}$ is expressed by the spin-dependent hopping term $t_s$\cite{Kregar2019}, 
\begin{equation}
\label{eq:ts}
t_s= t_0 e^{i \varphi_0 \left( \frac{1}{2} - \phi_m - s\phi_\alpha  \right)}.
\end{equation}
The phase shift, associated with electron's hopping, depends on both magnetic flux and the Rashba coupling. The magnetic flux is written in dimensionless form $\phi_m = {\pi R^2 B}/{\phi_0}$, with effective ring radius $R$, magnetic field $B$ and magnetic flux quantum $\phi_0$. The dimensionless form of the Rashba coupling $\alpha_R$ is written as $\alpha = {2 m R \alpha_R}/{\hbar}$. The Coulomb repulsion between electrons is expressed by matrix term $U_{n_1,n_2} = {e^2}/{[8\pi \epsilon \epsilon_0 \sin{ \left( \frac{\varphi_0}{2}\left|n_1-n_2 \right| \right)} ]}$ and counting operator $n_n =\sum_s d_{n s}^\dagger d_{n s}$.

The pseudo-spin diagonal form of Hamiltonian (\ref{eq:6_1_H0}) is a consequence of specific spin properties of operators $d^\dagger_{n s}$, explained in detail in Ref. \refcite{Kregar2019}, with their spin expectation values following the changes in orientation of the Rashba spin rotation axis along the ring. The expectation values of spin of electron in a state $ \left| \phi_{ns}(\varphi) \right\rangle =d^\dagger_{n s} \left| 0\right\rangle $ are therefore\cite{Kregar2019}
\begin{equation}
\label{eq:vecs}
\left\langle \vec{s} \right\rangle_{ns} = \frac{\hbar}{2} \left\langle\phi_{ns}(\varphi) \right| \vec{\sigma} \left| \phi_{ns}(\varphi) \right\rangle  = \hbar s \left[ \sin \vartheta_\alpha \cos \left( n \varphi_a \right) ,\sin \vartheta_\alpha \sin \left( n \varphi_a \right) , \cos \vartheta_\alpha \right].
\end{equation}
where the rotation angle $\theta_\alpha$ is defined as $\tan \theta_\alpha = -\alpha$ and $\varphi_0 = \frac{\pi}{2}$ is the angle distance between sites. 

\begin{figure}[htbp]
\centering
\includegraphics[scale=1]{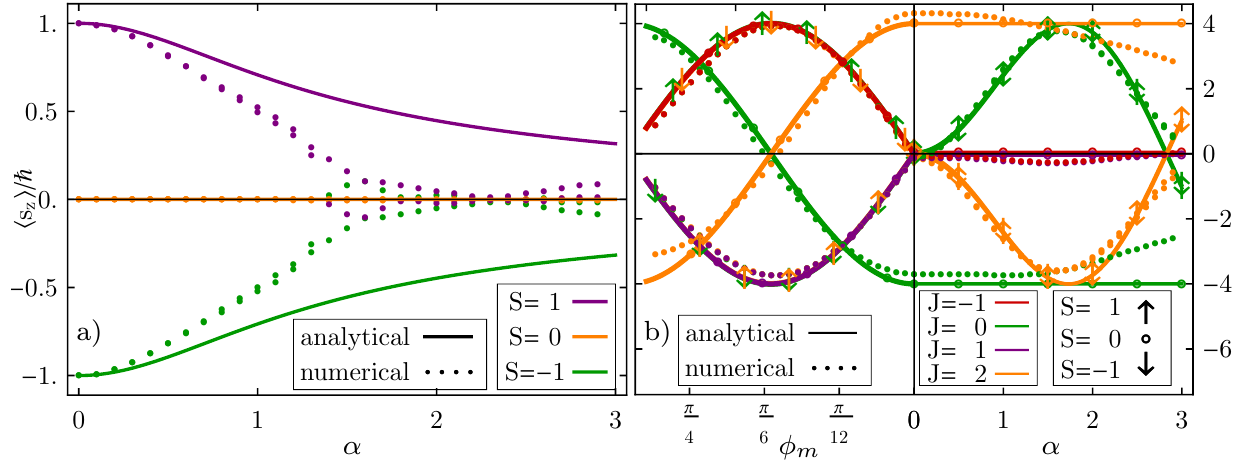}
\caption{Numerically calculated properties of two-electrons states in square quantum dot of size $L=800\,$nm (dots), compared to the results of analytic mapping to 4-site ring (solid lines). a) Expectation values of $z$ component os spin, $\left\langle s_z \right\rangle$. b) Eigenenergies of 8 low-energy states, shifted by average energy of all 8 states and renormalized to the magnitude of hopping term $\tilde{t}$.}
\label{fig:fig2}
\end{figure}

Since the total pseudo-spin $S=s_1 + s_2$ is conserved by the Hamiltonian (\ref{eq:6_1_H0}), the two-electron state of the system can be written as a superposition of two-electron basis states
\begin{equation}
\label{eq:2_3_basis}
\left| m n S \right\rangle = d^\dagger_{n s_1} d^\dagger_{n+m,s_2} \left| 0 \right\rangle,
\end{equation}
with $n$ determining the position of first electron and $m$ it's relative distance to the second electron. The conservation of $S$ allows the expectation value of $z$ component of spin to be determined without much effort from Eq. (\ref{eq:vecs}):
\begin{equation}
\label{eq:SzExpec}
\left\langle s_z \right\rangle = \hbar S \cos \vartheta_\alpha.
\end{equation}
This value is compared to numerically calculated expectation values for square two-electron quantum dot in Fig. \ref{fig:fig2} a). For small $\alpha$, the matching is very good, and even for larger values, the general trend of decreasing $\left\langle s_z \right\rangle$ is compatible with the numerical result. The fact that the numerically calculated magnitude of $\left\langle s_z \right\rangle$ decreases faster than predicted could probably be explained by relation between $\vartheta_\alpha$ and $\alpha$ in a square dot system being different from $\tan \vartheta = -\alpha$, but this effect was not studied further.

The total number of the basis states $\left| m n S \right\rangle$ of two electrons on $N=4$ sites is 28: $2 \times \left( 
\begin{array}{c}
4\\
2
\end{array}
\right) = 2 \times 6$ states for $S = \pm \frac{1}{2}$, and $4\times4 = 16$ states for $S=0$. Based on the value of Coulomb interaction $V_{m=n_1-n_2} = U_{n_1,n_2}$ between electrons in specific state, the basis states can be split into three subspaces: $V_{0}$ for states with both electrons on the same site ($m=0$), which can only occur for $S=0$, $V_{1}$ for electrons on neighbouring sites ($m=1,3$) and $V_2$ for electrons in the opposite corners of the dot ($m=2$). 
Full $28 \times 28$ Hamiltonian can therefore be written in block form 
\begin{align}
\mathcal{H}_{kin} + \mathcal{H}_C =& \left(
\begin{array}{c|c|c}
V_0 \times \mathbb{1}_{4\times 4} & \mathcal{H}_{kin,0\leftrightarrow1} & 0 \\ \hline
\mathcal{H}_{kin,0\leftrightarrow1}^\dagger & V_1 \times \mathbb{1}_{16\times 16}& \mathcal{H}_{kin,1\leftrightarrow2}\\ \hline
0 & \mathcal{H}_{kin,1\leftrightarrow2}^\dagger & V_2 \times \mathbb{1}_{8\times 8}
\end{array}
\right)
\end{align}

It is obvious that for a Coulomb interaction it holds $V_0 > V_1 >V_2$ and therefore the state with electrons on the opposite sites of the ring will be energetically preferable. The energy difference between states with a different charge configuration ($\Delta = V_1 - V_2$ and $\tilde{\Delta} = V_0 - V_2$) due to the Coulomb coupling increases with the size of the dot. In the limit of large $L$, where the Wigner molecule is formed, both differences are much larger then the magnitude of hopping term $t_s$ (\ref{eq:ts}): $\Delta,\tilde{\Delta} \gg \left| t_s \right| $.

The effective low-energy ring model will comprise only 8 basis states $\left| m n S \right\rangle$ with electrons in the low-energy configuration with $m=2$ and Coulomb energy $V_2$. 
Since different states in this subspace couple only via the states of $m=1$ subspace, the effective Hamiltonian is obtained by the second order perturbation theory for degenerate states. The energy gap of the order of $\Delta = V_1 - V_2$ between subspaces will result in effective hopping terms of magnitude $\sim \frac{t_s^2}{\Delta}$. The easiest way to construct the Hamiltonian is to use L\" owdin partitioning\cite{Lowdin1951}. The contribution of $m=0$ subspace, coupled only to the $m=1$ but not $m=2$ subspace, can be neglected, as it only produces effective terms in the 4th order of perturbation. The L\"owdin partitioning, performed on the remaining $24 \times 24$ matrix, results in the effective $8 \times 8$ Hamiltonian for subspace $V_2$:
\begin{equation}
\label{eq:6_H_eff}
\mathcal{H}_{8\times8} = \mathcal{H}_{0,8\times 8} + \mathcal{H}_{kin,8\times 8} = \left( - 4 \tilde{t} + V_2 \right) \mathbb{1}_{8\times 8}+ \left(
\begin{array}{c | c | c}
\mathcal{H}_{S=1} & 0 & 0 \\ \hline
0 & \mathcal{H}_{S=0} & 0 \\ \hline
0 & 0& \mathcal{H}_{S=-1} 
\end{array}
\right)
\begin{array}{c}
\left.\right\rbrace 2\\
\left.\right\rbrace 4\\
\left.\right\rbrace 2
\end{array}
\end{equation}
The first part of Hamiltonian (\ref{eq:6_H_eff}) is the constant term with the magnitude of effective hopping $\tilde{t} = \frac{t_0^2}{\Delta}$. The matrix part describes an effective coupling between states, which are now, with fixed $m=2$, labeled only by two quantum numbers, $n$ describing the orientation of the state and $S$ describing pseudo-spin.
\begin{equation}
\label{eq:2_3_states_nSd}
\left| n S \right\rangle = \left| m=2, n, S \right\rangle =  d^\dagger_{n, s_1} d^\dagger_{n+2, s_2} \left| 0 \right\rangle
\end{equation}
The non-diagonal matrix elements of Hamiltonian in this basis, $\mathcal{H}_{kin,8\times 8} = \sum_{mn}H_{Smn} \left| n S \right\rangle \left\langle m S \right| $ are
\begin{align}
\label{eq:HamS}
S=0: \quad H_{Smn} &= -2 \tilde{t} \left|f_{mn}\right| \exp( -i f_{mn} \left(2 \phi _m+1\right) ) \nonumber \\
S=\pm 1: \quad H_{Smn} &= -2 \tilde{t} f_{mn}  \cos \left[\varphi_0  \left(2 \phi_m + S \phi_\alpha \right)\right],
\end{align}
where $f_{mn} = \sin\left[\pi (m-n)/2 \right]$.
This effective hamiltonian can be understood as describing a hopping of electron pair from one dot diagonal to the other with the hopping term $2 \tilde{t}$. The hopping is accompanied by the acquisition of Peierls phase $2 \phi_m$, with factor 2 indicating that 2 electrons are involved in the process. Additionally, the pseudo-spin dependent phase $S \phi_\alpha = S \sqrt{1+\alpha^2}$ is acquired by the states with $S=\pm 1$.

Since hopping only occurs between neighbouring states, the eigenstates of Hamiltonian $\mathcal{H}_{8 \times 8 }$ are obtained by constructing rotating states from the basis states $\left| n S \right\rangle$ with pseudo-spin $S$ and total angular momentum $J$:
\begin{equation}
\label{eq:2_3_psi0}
\left| J S \right\rangle = \sum_{n} e^{i (J-1) n \varphi_0} \left| n S \right\rangle.
\end{equation}
Note that due to Pauli exclusion principle, only two values of angular momentum $J=0,2$ are compatible with $S=\pm 1$, while any value $J=0,1,2,3$ is compatible with $S=0$.

The Hamiltonian $\mathcal{H}_{8 \times 8 }$ is diagonal in basis $\left| J S \right\rangle$ with diagonal elements representing energy
\begin{equation}
\label{eq:2_3_eng}
E_{J S} = - 4 \tilde{t} \cos \left[ \varphi_0 \left( J - 2\phi_m -S \phi_\alpha \right) \right].
\end{equation}
This result is plotted alongside numerically calculated values in Fig. \ref{fig:fig2} b) as a function of both the Rashba coupling $\alpha$ and magnetic flux $\phi_m$. Note that the effective hopping term $\tilde{t}$ was here used as a fitting parameters used to map the results of ring model to the original square-quantum dot. Good agreement between results of both models can be observed $\alpha \lesssim 2$ and  $\phi_m \lesssim \pi/4$, indicating the plausibility of the mapping.

\section{Effective Hamiltonian in the Bell basis and electrically controlled qubit transformations}
The spin filtering device, proposed by Bayat\cite{Bayat2013}, makes use of magnetic flux controlled oscillations of singlet and triplet two-electron states in a square quantum dot. In the proposed effective 4-site model, this phenomenon can be explained by the vanishing of hopping terms in an effective Hamiltonian (\ref{eq:HamS}) at the Rashba coupling $\alpha = 0$ and magnetic flux parameters $\phi_m$ either integer or half-integer.

In the case of non-vanishing the Rashba coupling, the system dynamics become much more complex. Not only does the value of $\alpha$ affect the hopping magnitude of states in $S=\pm 1$ subspace via the Rashba coupling, but the basis states $\left| n S \right\rangle$ themselves also depend on $\alpha$ since the spin of states, created by operator $d_{n s}^\dagger$, is determined by $\alpha$. This suggests that the inclusion of the Rashba coupling will result in much richer possibilities of charge oscillation and spin rotation control. Note that the the Rashba coupling can be tuned by external electric field perpendicular to the plane of the system\cite{Datta1990}, allowing the external control of its strength.

To observe the effects of the Rashba coupling on the evolution of spin-charge states in the dot, it is convenient to rewrite the Hamiltonian from spin-dependent basis  $\left| n S \right\rangle$ into the basis of pure spin states,
\begin{equation}
\left| n S \right\rangle _c = c^\dagger_{n,s_1} c^\dagger_{n+2,s_2} \left| 0\right\rangle.
\end{equation}
where $c^\dagger_{n,s}$ is a standard creation operator for electron with spin either $s= \frac{1}{2} = \uparrow$ or $s= -\frac{1}{2} = \downarrow$. Operators $d^\dagger_{n,s}$ can be written as a linear combination of $s=\pm\frac{1}{2}$ operators $c^\dagger_{n,s}$ with same $n$, meaning that both sets of operators create electrons on the same site with only it's spin being rotated\cite{Kregar2019}:  
\begin{align}
d^\dagger_{n,\frac{1}{2}} &= e^{-i n \varphi_0} \cos{\left(\vartheta_\alpha/2\right)} c^\dagger_{n,\uparrow}  -  \sin{\left(\vartheta_\alpha/2\right)} c^\dagger_{n,\downarrow} \\
d^\dagger_{n,-\frac{1}{2}} &= e^{-i n \varphi_0} \sin{\left(\vartheta_\alpha/2\right)} c^\dagger_{n,\uparrow}  +\cos{\left(\vartheta_\alpha/2\right)} c^\dagger_{n,\downarrow}
\end{align}

The Bell states are defined as a specific combination of pure spin states. We introduce the basis of two sets of Bell states, one for each dot diagonal ($n = 0, 1$), defined as\cite{Nielsen2011}:
\begin{align}
\left| T_x, n \right\rangle \equiv & \left( \left| n,1 \right\rangle_c - \left| n,-1 \right\rangle_c  \right) /\sqrt{2} \sim \left( \left| \uparrow \uparrow \right\rangle - \left|\downarrow \downarrow \right\rangle \right)/\sqrt{2} \nonumber\\
\left| T_y, n \right\rangle \equiv& \left( \left| n, 1 \right\rangle_c  + \left| n,-1 \right\rangle_c  \right)/\sqrt{2} \sim \left( \left| \uparrow \uparrow \right\rangle + \left|\downarrow \downarrow \right\rangle \right)/\sqrt{2} \\
\left| T_z, n \right\rangle \equiv&  \left( \left| n, 0 \right\rangle_c  + \left| n+2,0 \right\rangle_c  \right)/\sqrt{2} \sim \left( \left| \uparrow \downarrow \right\rangle + \left|\downarrow \uparrow \right\rangle \right)/\sqrt{2} \nonumber \\
\left| S_{\phantom{0}}, n \right\rangle \equiv& \left( \left| n, 0 \right\rangle_c  - \left| n+2,0 \right\rangle_c  \right)/\sqrt{2} \sim \left( \left| \uparrow \downarrow \right\rangle - \left|\downarrow \uparrow \right\rangle \right)/\sqrt{2}\nonumber,
\end{align}
with triplet states denoted $\left| T_i, n \right\rangle$ and singlet $\left| S, n \right\rangle$.

The effective Hamiltonian $\mathcal{H}_{kin}$ (\ref{eq:HamS}) in the basis of Bell states can be expressed by matrix
\begin{equation}
\label{eq:HbellTrans}
\mathcal{H}_{kin,Bell} = \left( \begin{array}{c|c}
0 & \mathcal{H}_{0 \leftrightarrow1 } \\ \hline
\mathcal{H}^\dagger_{0 \leftrightarrow1 } & 0\\
\end{array} \right);
\mathcal{H}_{0 \leftrightarrow1 } =\left(
\begin{array}{cccc}
-4 i s_{\phi } t_2 & \phantom{-}4 i c_{\phi } t_5 & 0 &\phantom{-}4 s_{\phi } t_1 \\
\phantom{-}4 i c_{\phi } t_3 & -4 i s_{\phi } t_2 & 0 &\phantom{-}4 c_{\phi } t_6 \\
0 & 0 &  -4 i\tilde{t}  s_{\phi } & 0 \\
\phantom{-}4 i c_{\phi } t_6 &\phantom{-}4 i s_{\phi } t_1 & 0 &\phantom{-}4 c_{\phi } t_4 \\ 
\end{array}
\right).
\end{equation}
The kinetic Hamiltonian couples only the Bell states with different orientation $n$, therefore only off-diagonal blocks of matrix are non-vanishing. The matrix is relatively sparse, with coefficients being
\begin{equation}
\begin{array}{lll}
s_\phi = \sin \left(\pi \phi_m \right)
&\phantom{x}c_\phi = \cos \left(\pi \phi_m \right) 
&\phantom{x}s_\alpha = \sin \vartheta_\alpha \quad  c_\alpha = \cos \vartheta_\alpha\\
t_1 = \tilde{t} s_\alpha \sin \left( \frac{\pi}{2} \phi_\alpha \right) 
&\phantom{x}t_2 = \tilde{t} c_\alpha \sin \left( \frac{\pi}{2} \phi_\alpha \right) 
&\phantom{x}t_3 = \tilde{t} \left[ s_\alpha^2 + \cos \left( \frac{\pi}{2} \phi_\alpha \right) c_\alpha^2 \right]\\
t_4 = \tilde{t} \left[ c_\alpha^2 + \cos \left( \frac{\pi}{2} \phi_\alpha \right) s_\alpha^2 \right] 
&\phantom{x}t_5 = \tilde{t} \cos \left( \frac{\pi}{2} \phi_\alpha \right) 
&\phantom{x}t_6 = \tilde{t} s_\alpha c_\alpha \left[ 1 - \cos \left( \frac{\pi}{2} \phi_\alpha \right)\right] 
\end{array}
\end{equation}
We see that all hopping terms are proportional to the effective hopping $\tilde{t}$, while their phases and magnitudes are affected by magnetic flux and the Rashba coupling. When $\alpha = 0$, all off-diagonal terms in $\mathcal{H}_{Bell, 0 \leftrightarrow1 }$, coupling singlet and triplet states, vanish. In this case, studied by Bayat\cite{Bayat2013}, the charge oscillations are controlled by magnetic flux in terms $s_\phi$ and $c_\phi$, distinguishing singlet and triplet states.

When the Rashba coupling is present, several off-diagonal term in (\ref{eq:HbellTrans}) are coupling singlet and triplet states, resulting in their mixing during charge oscillations. The mixing is especially simple if the magnetic flux is set to 
$\phi_m = 1/2$ ($c_\phi = 1$, $s_\phi = 0$). In this case, the $8\times8$ Hamiltonian (\ref{eq:HbellTrans}) can be split into three subspaces that do not mix with each other, spanned by the basis, 
$\Psi_{T_x} = \left\lbrace \left| T_x, 1 \right\rangle ,\left| S, 1 \right\rangle ,\left| T_x,0 \right\rangle  \right\rbrace$,
$\Psi_{T_y} = \left\lbrace \left| T_y, 0 \right\rangle ,\left| S, 0 \right\rangle,\left| T_y, 1 \right\rangle \right\rbrace$,
$\Psi_{T_z} = \left\lbrace \left| T_z, 0 \right\rangle, \left| T_z, 1 \right\rangle \right\rbrace$. The Hamiltonian of last subspace $\Psi_{T_z}$ is especially simple since its off-diagonal coupling terms are independent of either $\alpha$ or $\phi_m$. The Hamiltonians for other two subspaces, coupling singlet and triplet state on one diagonal with tripet state on the other diagonal, are very similar and we will therefore focus the analysis on a single subspace $\Psi_{T_y}$. 

\begin{figure}[htbp]
\includegraphics[scale=1]{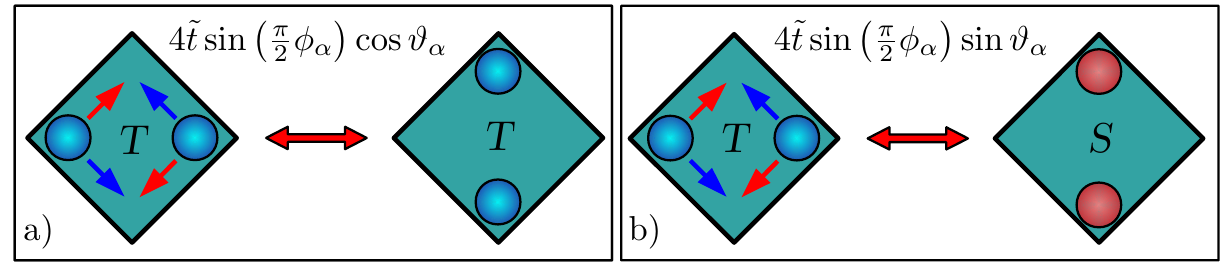}
\caption{Schematic representation of hopping terms between Bell states in a square quantum dot. The hopping part of the Hamiltonian only couples the states with different charge distribution. Triplet-triplet coupling (Fig.~a)) is proportional to $\cos \vartheta_\alpha$, while triplet-singlet coupling (Fig.~b)) is proportional to $\sin \vartheta_\alpha$.}
\label{fig:fig3}
\end{figure}

The charge oscillations of states in this subspace are governed by Hamiltonian
\begin{equation}
\mathcal{H}_{Bell,T_y} = - i t_\alpha \left( c_\alpha \left| T_y, 0 \right\rangle + s_\alpha\left| S, 0 \right\rangle  \right) \left\langle T_y, 1 \right| + {\rm h.c.}
\end{equation}
where $t_\alpha = 4 \tilde{t} \sin \left( \frac{\pi}{2} \phi_\alpha \right)$. Singlet and triplet states on the same diagonal do not interact directly, but only via the third state, triplet, positioned on the other diagonal, as shown schematically in Fig.~\ref{fig:fig3}.

The total hopping magnitude between diagonals 
\begin{equation}
\left| i t_\alpha c_\alpha \right|^2 +\left| - i t_\alpha s_\alpha \right|^2  = t_\alpha^2 = 16 \tilde{t}^2 \sin^2 \left( \frac{\pi}{2} \phi_\alpha \right)
\end{equation}
depends on the Rashba coupling $\alpha$, allowing for electric control of oscillation rate, based on spin symmetry properties of the two-electron states. What is even more important is that the ratio between hoping amplitudes for singlet and triplet state also depends on $\alpha$:
\begin{equation}
\frac{\mathcal{H}_{\left| S,_{\phantom{y}} 0 \right\rangle \rightarrow \left| T_y, 1 \right\rangle}}{\mathcal{H}_{\left| T_y, 0 \right\rangle \rightarrow \left| T_y, 1 \right\rangle}} = \frac{- i t_\alpha s_\alpha}{\phantom{-}i t_\alpha c_\alpha} = -\tan \vartheta_\alpha
\end{equation}
This indicates that the magnitude of the Rashba coupling can be used to control the rotation from singlet to triplet state and vice versa during the oscillation of charge between dot diagonals. In the qubit system with states based on singlet and triplet Bell states, this would allow for a controlled transition between states based on the external tuning of the Rashba interaction.

\section{Conclusion}

In the present paper, the system of two-electrons in a mesoscopic square quantum dot in the presence of the Rashba coupling was studied theoretically with the focus on its potential application for quantum data processing. The problem was simplified by mapping the Hamiltonian of the quantum dot to a system of two electrons on a 4-site quantum ring by comparing the charge distribution, energy and spin of eigenstates of both systems. The resulting $8\times8$ effective Hamiltonian allowed the analysis of charge oscillations in the presence of external magnetic flux, reproducing the already known properties of the studied system, which further confirmed the validity of the mapping.

The Hamiltonian terms in the presence of the Rashba coupling show that the charge oscillation frequency could also be tuned by externally controlled Rashba coupling strength, providing additional control over the system evolution. Even more importantly, the mixing between singlet and triplet states in the dot, induced by charge oscillation,  can be controlled by the Rashba coupling, which opens new possibilities for fully electric control of spin-based qubit states.

The properties of the system should in the future be further explored by studying in detail the time dynamics of the system evolution. The control over the system could potentially be further enhanced by the addition of external voltage gate, controlling the electrostatic energy of different charge configuration. Two-qubit gates, realized via electrostatic interaction between neighbouring dots, could also be explored into more detail, potentially leading to a qubit system fulfilling all five DiVincenzo criteria for a qubit needed to construct a universal quantum computer\cite{DiVincenzo2000}.


\end{document}